\documentclass{PoS}

\usepackage{amsmath}

\title{Charm physics prospects at Belle~II}

\ShortTitle{Charm physics prospects at Belle~II}

\author{\speaker{Giacomo De Pietro}%
        \thanks{On behalf of the Belle~II Collaboration.}\\
       Dipartimento di Matematica e Fisica, Universit\`a di Roma Tre and INFN Sezione di Roma Tre,\\Via della vasca navale 84, I-00146 Rome, Italy\\
       E-mail: \email{giacomo.depietro@roma3.infn.it}}

\abstract{Belle~II is a major upgrade of the Belle experiment and will operate at the $B$-factory SuperKEKB in Japan. Here we discuss the expected sensitivity of Belle~II for $D^0 - \bar{D}^0$ mixing and $CP$ violation measurements in the charm sector, which will benefit from a factor 50 increase in statistics and an improved vertex detection and particle identification. The impact on the determination of CKM parameters from the measurements of purely leptonic $D$ mesons decays is discussed. Finally a novel method of flavour tagging to substantially increase the sample of $D^0$ and $\bar{D}^0$ is presented.}

\FullConference{EPS-HEP 2017, European Physical Society conference on High Energy Physics\\
		5-12 July 2017\\
		Venice, Italy}

\begin{document}

\section{The Belle~II experiment}
Belle~II is a major upgrade of the Belle experiment~\cite{tdr} and will operate at the $B$-factory SuperKEKB, located at the KEK laboratory in Tsukuba, Japan. Although Belle~II has been designed to perform precise measurements in the $b$-quark sector, it will also be an ideal laboratory to study the properties of the charm quark. The data taking will start in 2018 and Belle~II is expected to collect within the next decade a data sample of more than $10^{10}$ $c\bar{c}$ events with a total integrated luminosity of about 50 ab$^{-1}$.

Given the clean environment of the $e^+ e^-$ SuperKEKB collider and the hermiticity of the detector, Belle~II is expected to have great perfomances in the reconstruction of final states with neutral particles (e.g. $\gamma$, $\pi^0$, $\eta$) and missing energy (e.g. leptonic and semileptonic decays of $D$ mesons).

Moreover, Belle~II will have a six-layer silicon vertex detector (2 layers of DEPFET pixel detectors and 4 layers of double-sided silicon strip detectors) whose innermost layer will be 2 times closer to the interaction point with respect to Belle. From Monte Carlo (MC) simulations it has been shown that for the processes $D^0 \to h^- h^+$ (where $h = \pi, K$)
Belle~II will have a $D^0$ proper time resolution $\sigma_t = $~0.14~ps~\cite{b2tip}, which is about half with respect to BaBar (0.27~ps).

\section{Impact on time-dependent $D^0 - \bar{D}^0$ mixing and $CP$ violation measurements}

One of the main goals of the Belle~II charm physics program is to improve the current measurements of $D^0 - \bar{D}^0$ mixing and the search for the $CP$ violation (CPV) in the charm sector.

The time-dependent analysis of the ratio $R_{WS}$ of the so-called ``wrong sign'' (WS) decay $D^0 \to K^+~\pi^-$ to the ``right sign'' (RS) decay $D^0 \to K^-~\pi^+$\footnote{Throughout this proceeding, charge-conjugate modes are implicitly included.} is sensitive to both the mixing parameters ($x^{\prime} = x \cos\delta_{K\pi} + y\sin\delta_{K\pi}$, $y^{\prime} = y \cos\delta_{K\pi} - x\sin\delta_{K\pi}$, where $x=\Delta m / \Gamma$, $y = \Delta \Gamma / 2\Gamma$ and $\delta_{K\pi}$ is the strong phase) and the $CP$-violating parameters ($|q/p|$ and $\phi = \arg(q/p)$):
\begin{equation}
R_{WS} = R_D + \left| \frac{q}{p} \right| \sqrt{R_D} (y^{\prime} \cos\phi - x^{\prime} \sin\phi)(\Gamma t) \\+ \left| \frac{q}{p} \right|^2 \frac{x^{\prime 2} + y^{\prime 2}}{4} (\Gamma t)^2 \mbox{ ,} \label{eq:kpi}
\end{equation}
where $R_D = |\mathcal{A}(D^0~\to K^+~\pi^-) / \mathcal{A}(D^0~\to K^-~\pi^+)|^2$.

An ensemble of 1000 toy MC experiments, including the smearing of the decay times by the 0.14~ps expected proper time resolution, was generated to test the sensitivity of Belle~II with the full data set~\cite{b2tip}.
The preliminary estimates of the expected sensitivity with 50 ab$^{-1}$ of data are:
$\sigma_{x^{\prime}} = 0.15\%$, $\sigma_{y^{\prime}} = 0.10\%$,
$\sigma_{|q/p|} = 0.05\%$ and $\sigma_{\phi} = 5.7^{\circ}$.

The time-dependent decay rate of the WS decay $D^0 \to K^+~\pi^-~\pi^0$ has been also studied (in this phase of the study, possible $CP$ violation and backgrounds have been neglected). The decay rate for the WS decays at a given point in the Dalitz plot $(s_{12},s_{13})=(m^2_{K^+ \pi^-},m^2_{K^+ \pi^0})$ may be written as:
\begin{equation}
\frac{d\Gamma(D^0 \to K^+\pi^-\pi^0)}{dt\,ds_{12}\,ds_{13}} \propto r_0^2 \left|\mathcal{A}^{\rm DCS}_{\bar{f}}\right|^2 + r_0 \Big(A y^{\prime\prime}  + B x^{\prime\prime}\Big)(\Gamma t) + \frac{(x^{\prime\prime 2} + y^{\prime\prime 2})}{4} \left|\bar{\mathcal{A}}^{\rm CF}_{\overline{f}}\right|^2(\Gamma t)^2 \mbox{ ,} \label{eq:kpipi0}
\end{equation}
where $x^{\prime\prime}$ and $y^{\prime\prime}$ are the mixing parameters rotated by the strong phase $\delta_{K\pi\pi^0}$ and $A$, $B$ and $r_0$ are terms depending only on the decay amplitudes (in Equation~\ref{eq:kpipi0} $|q/p|=1$ and $\phi=0$ since $CP$ violation has been neglected).

An ensemble of 10 experiments, each consisting of $\sim 2.2\cdot 10^5$ signal events corresponding to the full Belle~II dataset (assuming similar efficiency to BaBar)~\cite{kpipi0}, was generated for this study, taking into account the expected proper time resolution. The estimates of the expected sensitivity are: $\sigma_{x} = 0.049\%$ and $\sigma_{y} = 0.057\%$.

These results well represent a significant improvement with respect to the Belle and BaBar achievements~\cite{hfag}.

\section{Impact on time-integrated $CP$ violation measurements}

Belle~II will have excellent efficiency for reconstructing multi-body final states with tiny and well-controlled detector-based asymmetries. Thus the experiment is ideal for searching for time-integrated $CPV$ in many final states. The expected precision of Belle~II for the measurement of a $CP$ asymmetry $A_{CP} = [\Gamma(D \to X) - \Gamma(\bar{D} \to \bar{X})]/[\Gamma(D \to X) + \Gamma(\bar{D} \to \bar{X})]$ is obtained by properly scaling the Belle uncertainty by the integrated luminosity~\cite{b2tip}.

In particular, if $\sigma_{\text{stat}}$ is the statistical error of the Belle measurements, $\sigma_{\text{syst}}$ is the systematic error that scales with the luminosity (e.g. background shapes measured with control samples) and $\sigma_{\text{irred}}$ is the systematic error that doesn't scale (e.g. vertexing and tracking resolution due to the detector misalignement), the expected uncertainty of Belle~II $\sigma_{\text{Belle~II}}$ with the full integrated luminosity is given by:
\begin{equation} 
\sigma_{\text{Belle~II}} = \sqrt{(\sigma^2_{\text{stat}} + \sigma^2_{\text{syst}})\cdot (\mathcal{L}_{\text{Belle}}/50~\text{ab}^{-1}) + \sigma^2_{\text{irred}}} \mbox{ .} \label{eq:error}
\end{equation}
A list of $D^0$, $D^+$ and $D_s^+$ modes for which Belle has measured $A_{CP}$ and the expected uncertainty of Belle~II $\sigma_{\text{Belle~II}}$ is given in Table~\ref{tab:cpv_list}.
\begin{table} 
  \centerline{
    \begin{tabular}{l|cc|c} 
	\hline \hline
	Channel                            & \multicolumn{2}{c|}{Belle measurement}  & $\sigma_{\text{Belle~II}}$ \\
		                           & $\mathcal{L}\:(\mbox{fb}^{-1})$ & $A_{CP}\:(\%)$   & 50 $\mbox{ab}^{-1}$ \\
	\hline \hline
	$D^0\to K^+K^-$                    & 976 & $-0.32\,\pm 0.21\,\pm 0.09$        & $\pm 0.06$ \\
	$D^0\to \pi^+\pi^-$                & 976 & $+0.55\,\pm 0.36\,\pm 0.09$        & $\pm 0.06$ \\
	$D^0\to \pi^0\pi^0$                & 966 & $-0.03\,\pm 0.64\,\pm 0.10$        & $\pm 0.09$ \\
	$D^0\to K_S^0\,\pi^0$              & 966 & $-0.21\,\pm 0.16\,\pm 0.07$        & $\pm 0.03$ \\
	$D^0\to K_S^0\,\eta$               & 791 & $+0.54\,\pm 0.51\,\pm 0.16$        & $\pm 0.07$ \\
	$D^0\to K_S^0\,\eta'$              & 791 & $+0.98\,\pm 0.67\,\pm 0.14$        & $\pm 0.09$ \\
	$D^0\to K_S^0\,K_S^0$              & 921 & $-0.02\,\pm 1.53\,\pm 0.17$	      & $\pm 0.21$ \\
	$D^0\to K^+\pi^-\pi^0$             & 281 & $-0.60\,\pm 5.30$                  & $\pm 0.40$ \\
	$D^0\to K^+\pi^-\pi^+\pi^-$        & 281 & $-1.80\,\pm 4.40$                  & $\pm 0.33$ \\
	$D^0\to \pi^+\pi^-\pi^0$           & 532 & $+0.43\,\pm 1.30$                  & $\pm 0.13$ \\
	$D^0\to \rho^0\,\gamma$            & 976 & $+0.056\,\pm 0.152\,\pm 0.006$     & $\pm 0.02$ \\  
	$D^0\to \phi\,\gamma$              & 976 & $-0.094\,\pm 0.066\,\pm 0.001$     & $\pm 0.01$ \\
	$D^0\to \overline{K}^{*0}\,\gamma$ & 976 & $-0.003\,\pm 0.020\,\pm 0.000$     & $\pm 0.003$ \\
	\hline
	$D^+\to \pi^0\pi^+$                & \multicolumn{2}{c|}{to be published}     & $\pm 0.40$ \\
	$D^+\to \phi\pi^+$                 & 955 & $+0.51\,\pm 0.28\,\pm 0.05$        & $\pm 0.04$ \\
	$D^+\to \eta\pi^+$                 & 791 & $+1.74\,\pm 1.13\,\pm 0.19$        & $\pm 0.14$ \\
	$D^+\to \eta'\pi^+$                & 791 & $-0.12\,\pm 1.12\,\pm 0.17$        & $\pm 0.14$ \\
	$D^+\to K^0_S\,\pi^+$              & 977 & $-0.36\,\pm 0.09\,\pm 0.07$        & $\pm 0.03$ \\
	$D^+\to K^0_S\,K^+$                & 977 & $-0.25\,\pm 0.28\,\pm 0.14$        & $\pm 0.05$ \\
	\hline
	$D^+_s\to K^0_S\,\pi^+$            & 673 & $+5.45\,\pm 2.50\,\pm 0.33$        & $\pm 0.29$ \\
	$D^+_s\to K^0_S\,K^+$              & 673 & $+0.12\,\pm 0.36\,\pm 0.22$        & $\pm 0.05$ \\
	\hline \hline
    \end{tabular}
  }
  \caption{A list of $A_{CP}$ measurement performed by Belle with the expected uncertainty for Belle~II (as defined in Equation~\protect\ref{eq:error}). For most of the decay modes considered the scaled uncertainty on $A_{CP}$ is less than 0.1$\%$.}
  \label{tab:cpv_list}
\end{table}

Two processes are specially interesting for Belle~II: $D^0 \to K^0_S K^0_S$ ($A_{CP}$ is enhanced up to 1\% in Standard Model (SM) predictions~\cite{ksks}) where we expect $\sigma_{\text{Belle~II}} = 0.21\%$, and $D^+ \to \pi^+ \pi^0$ (the SM predicts $A_{CP}=0$~\cite{pipi0}, so CPV could be enhanced by New Physics contributions) where we expect $\sigma_{\text{Belle~II}} = 0.40\%$.

\section{Impact on purely leptonic decays}

The clean environment of a $B$-factory and the knowledge of the center-of-mass energy will allow Belle~II to measure precisely the branching fractions of the leptonic decays $D_{q}^- \to l^- \bar{\nu}_l$ (with $q=d,s$) and to extract the quantities $f_{D_q} |V_{cq}|$ (where $f_{D_q}$ are the decay constants and $V_{cq}$ are the CKM matrix elements). By using the $f_{D_q}$ values computed with lattice QCD techniques, Belle~II will significantly improve the Belle measurement of $|V_{cs}|$ (that is the world's most precise determination so far) and will measure $|V_{cd}|$ with less than 2\% uncertainty~\cite{b2tip}.

The so-called ``charm tagger'' method was used in Belle to measure the charmed leptonic decays~\cite{zupanc} and it is foreseen be used also in Belle~II. It consists in the reconstruction of a ``tag side'' $D_{\rm tag}$ (hadronic decays of $D^0$, $D^+$ or $\Lambda_c^+$) recoiling against the signal $D_{q}^- \to l^- \bar{\nu}_l$. The remaining pions, kaons and protons are grouped together into the so-called ``fragmentation side'' $X_{\rm frag}$ (the fragmentation side is constructed imposing the conservation of the strangeness and baryonic numbers of the entire process). The missing momentum $P_{\rm miss} = P_{CM} - P_{\rm tag} - P_{\rm frag} - P_{l^-}$ is then constructed, and for the signal processes $D_{q}^- \to l^- \bar{\nu}_l$ the quantity $P^2_{\rm miss} \equiv m^2_{\nu}$ peaks at 0 GeV/c$^2$.

This methods will allow Belle~II also to improve the search for the invisibles decays of $D^0$ mesons (e.g. $D^0 \to \nu \bar{\nu}$).

\section{The ROE method for the flavour tagging of $D^0$ and $\bar{D}^0$}

\begin{figure}
  \centerline{
    \includegraphics[width=0.7\textwidth]{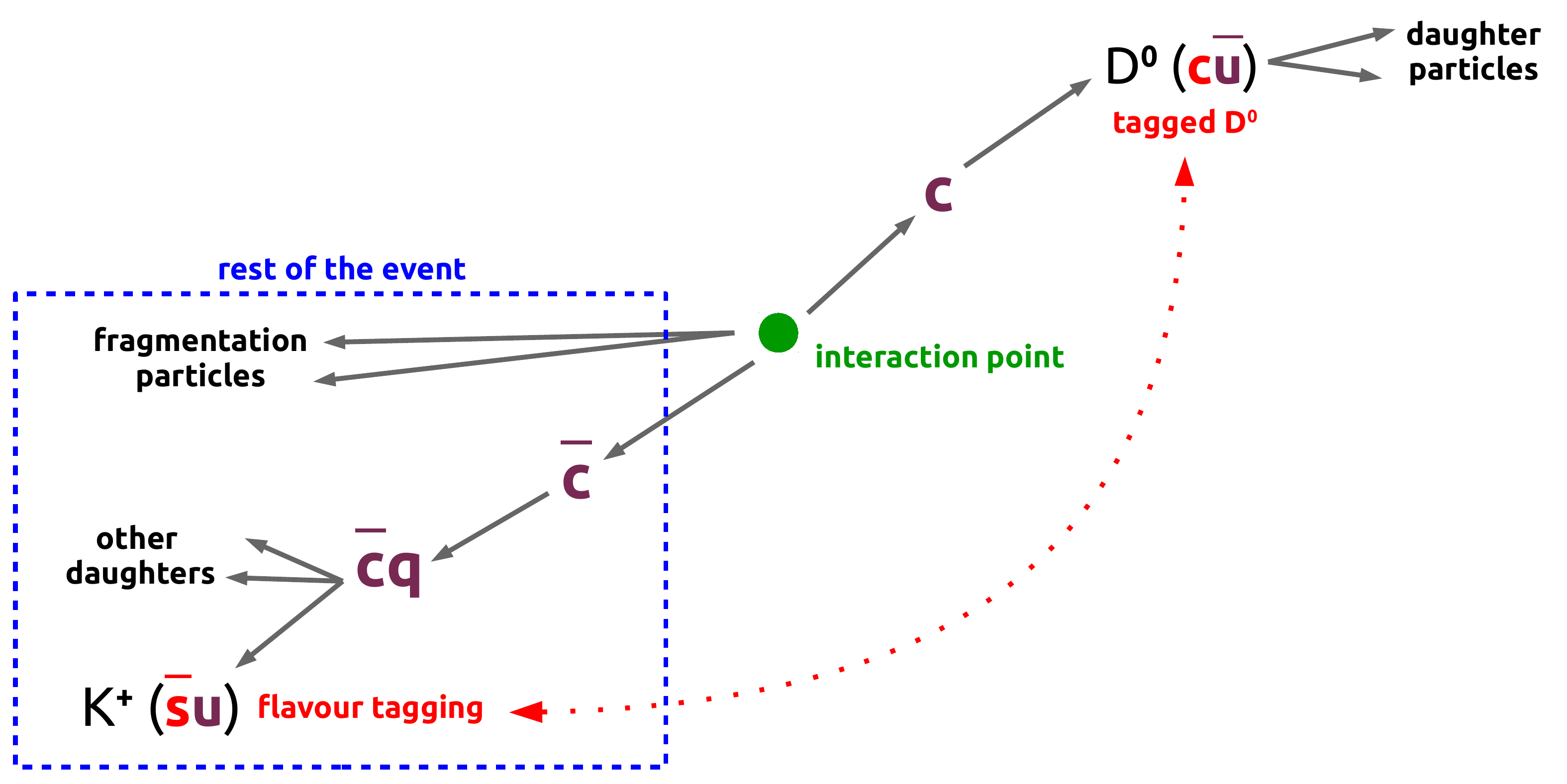}
  }
  \caption{The ROE method for the flavour tagging: only the events with a single $K^{\pm}$ candidate in the ROE are selected to tag the flavour of the neutral $D$ meson in a $c\bar{c}$ event.}
  \label{fig:roe}
\end{figure}

A new flavour tagging method, called ``ROE method'', has been developed in order to determine the flavour of $D^0$ and $\bar{D}^0$ at the production time in a $c\bar{c}$ event~\cite{b2tip}. This method, which is complementary to the one using the pion charge in the strong decay $D^{*+} \to D^0 \pi^+$ decay, is sketched in Figure~\ref{fig:roe} and exploits events where a signal $D^0$ is reconstructed and only one $K^{\pm}$ is identified in the rest of the event (ROE). In such case the charge of the kaon tags the flavour of the neutral $D$ meson (a $K^+$ in the ROE tags a $D^0$, instead a $K^-$ tags a $\bar{D}^0$).

A full simulation was performed using the MC data in order to study and evaluate the performances of this new flavour tagging technique. The expected performances are the following: tagging efficiency $\epsilon \sim 27\%$, mistagging fraction $\omega \sim 13\%$ and effective tagging efficiency $Q = \epsilon(1-2\omega) \sim 20\%$. When comparing such performance with the $D^{*+}$ tagging method, which yelds $Q \sim 80\%$, it should be taken into account that only 25\% of the $D^0$ mesons comes from a $D^{*+}$ decay in a $c\bar{c}$ event at the $B$-factories.

Since a fraction of $D^0$ can be doubly tagged using both methods, it will be possible to measure $\epsilon$ and $\omega$ for the ROE method on the Belle~II data, using the $D^{*+}$ method as reference. An integrated luminosity of $\sim 13$~fb$^{-1}$ will be enough to measure the mistagging of the ROE method with an uncertainty $\sigma_{\omega} \lesssim 1\%$.

It has been estimated that, by combining the results obtained using the $D^{*+}$ tag and the ROE method, a 15\% reduction of the statistical uncertainty on $A_{CP}(D^0 \to K^- \pi^+)$ can be achieved. We can interpret this reduction of the statistical uncertainty as having an ``effective'' luminosity higher by 35\% than the nominal one of SuperKEKB.

\section{Conclusions}

The Belle~II experiment will start to collect data in 2018 and its charm physics program will cover a large number of topics. Thanks to the large dataset (50~ab$^{-1}$) and the introduction of new experimental techniques it is expected to improve the $D^0 - \bar{D}^0$ mixing and $CP$ violation measurements performed at the $B$-factories, reaching the precision of the SM predictions for many final states. Also the measurement of the charmed leptonic decays will benefit the large amount of data that Belle~II will collect, improving the existing determination of the CKM matrix element $V_{cs}$ and $V_{cd}$.

To sum up, it is foreseen that Belle~II will be, in the next years, one of the top players in the charm sector measurements.

\end{document}